\documentclass[reprint, showpacs, preprintnumbers, aps, amsmath,amssymb, prb]{revtex4-1}
\usepackage{graphicx}
\usepackage{dcolumn}
\usepackage{bm}
\usepackage{color}
\usepackage{ulem}

\begin{document}


\title{Charge-spin-orbital fluctuations in
mixed valence spinels: \\
comparative study of AlV$_2$O$_4$ and LiV$_2$O$_4$}
\author{Amane Uehara,$^1$ Hiroshi Shinaoka,$^{2,3}$ and Yukitoshi Motome$^1$}
\affiliation{$^{1}$Department of Applied Physics, University of Tokyo, Tokyo 113-8656, Japan}
\affiliation{$^{2}$Theoretische Physik, ETH Z\"{u}rich, 8093 Z\"{u}rich, Switzerland}
\affiliation{$^{3}$Department of Physics, University of Fribourg, 1700 Fribourg, Switzerland}
\date{\today}

\begin{abstract}
Mixed valence spinels provide a fertile playground for the interplay between charge, spin, and orbital degrees of freedom in strongly correlated electrons on a geometrically frustrated lattice.
Among them, AlV$_2$O$_4$ and LiV$_2$O$_4$ exhibit contrasting and puzzling behavior: self-organization of seven-site clusters and heavy fermion behavior.
We theoretically perform a comparative study of charge-spin-orbital fluctuations in these two compounds, on the basis of the multiband Hubbard models constructed by using the maximally-localized Wannier functions obtained from the {\it ab initio} band calculations.
Performing the eigenmode analysis of the generalized susceptibility, we find that, in AlV$_2$O$_4$, the relevant fluctuation appears in the charge sector in $\sigma$-bonding type orbitals.
In contrast, in LiV$_2$O$_4$, optical-type spin fluctuations in the $a_{\rm 1g}$ orbital are enhanced at an incommensurate wave number at low temperature.
Implications from the comparative study are discussed for the contrasting behavior, including the metal-insulator transition under pressure in LiV$_2$O$_4$.
\end{abstract}


\maketitle

\section{Introduction\label{sec:introduction}}
Multiple degrees of freedom of electrons in solids, i.e., charge, spin, and orbital, have been one of the central issues in strongly correlated electron systems.
Their interplay is a source of various fascinating properties in magnetism, transport, and optics, which, in turn, provides the possibility of controlling the diverse quantum many-body phenomena~\cite{Tokura-Nagaosa}.
In particular, in the systems whose lattice structures are geometrically frustrated, such interplay becomes more conspicuous; charge-spin-orbital entangled fluctuations are promoted by keen competition between many different quantum states, and result in a broader range of peculiar behavior uniquely found in frustrated strongly correlated electrons~\cite{Introduction-to-Frustrated-Magnetism}.

Spinels, one of the primary minerals, provide a rich playground for studying such unique properties~\cite{S.Lee-2010-Spinel}.
Among them, a series of vanadium spinel oxides with mixed valence are of particular interest, as they exhibit a wide range of peculiar properties, from metal-insulator transition to superconductivity.
In the present study, we are particularly interested in two cousins in the vanadium spinel oxides, AlV$_2$O$_4$ and LiV$_2$O$_4$.
In both compounds, V cations comprise a pyrochlore lattice structure with strong geometrical frustration, and the threefold $t_{\rm 2g}$ orbitals are partially occupied by a half-integer number of $3d$ electrons: $(3d)^{2.5}$ in AlV$_2$O$_4$ and $(3d)^{1.5}$ in LiV$_2$O$_4$.
Despite the common aspects, two compounds show contrasting behavior, as described below.

AlV$_2$O$_4$ exhibits a structural phase transition at $T_{\rm c}=700$~K~\cite{K.Matsuno-AlV2O4-2001}.
Below $T_{\rm c}$, the pyrochlore lattice of V cations is distorted so as to self-organize seven-site clusters called ``heptamers"~\cite{Y.Horibe-AlV2O4-2006}.
The low-temperature ($T$) insulating state is interpreted by coexistence of a spin-singlet state in each heptamer and nearly free magnetic moments at the rest V sites~\cite{Y.Horibe-AlV2O4-2006,K.Matsuda-AlV2O4-2006}.
However, it remains unclear how the self-organization takes place from the interplay between charge, spin, and orbital degrees of freedom and the coupling to lattice distortions.

In contrast, LiV$_2$O$_4$ remains metallic down to the lowest $T$ without showing any symmetry breaking~\cite{Rogers-LiV2O4-1967}.
The compound, however, exhibits peculiar heavy fermion behavior at low $T$, which is unexpected in $3d$ electron systems~\cite{S.Kondo-LiV2O4-1997,C.Urano-LiV2O4-2000}.
Despite a number of intensive studies~\cite{V.Anisimov-LiV2O4-1999, V.Eyert-LiV2O4-1999, H.Kusunose-HF-2000, H.Tsunetsugu-LiV2O4-2002, Y.Yamashita-LiV2O4-2003, R.Arita-LiV2O4-2007, V.Yushankhai-LiV2O4-2007, K.Hattori-LiV2O4-2009}, the origin of this heavy fermion behavior remains elusive.
Interestingly, LiV$_2$O$_4$ exhibits a metal-insulator transition in an external pressure~\cite{C.Urano-LiV2O4-PHD-2000, K.Takeda-LiV2O4-2005, A.Irizawa-LiV2O4-2011}.
Although a structural change similar to AlV$_2$O$_4$ was suggested at this pressure-induced transition~\cite{P.Gaudart-LiV2O4-2007}, the relation between the two compounds and the nature of the transition have also remained as an unsolved issue.

In this study, for understanding of the contrasting behavior in these cousin compounds, we investigate the instability from the high-$T$ paramagnetic phase by analyzing fluctuations in the charge, spin, and orbital degrees of freedom.
For a multiband Hubbard model constructed on the basis of the {\it ab initio} band calculations and the analysis by means of the maximally-localized Wannier functions (MLWF),
we compute the generalized susceptibility, which describes the fluctuations in the charge, spin, and orbital channels, at the level of the random phase approximation (RPA).
We clarify the nature of relevant fluctuations by analyzing the eigenmodes of the generalized susceptibility.
We find that Coulomb interactions enhance dominantly $\sigma$-type charge fluctuations in AlV$_2$O$_4$, while optical-type spin fluctuations in LiV$_2$O$_4$.
We also discuss the pressure effect for adding theoretical inputs to the mechanism of pressure-induced metal-insulator transition in LiV$_2$O$_4$.

The organization of this paper is as follows.
In Sec.~\ref{sec:model-method}, we introduce the model and method.
After introducing the multiband Hubbard models constructed from MLWFs obtained by {\it ab initio} band calculations in Sec.~\ref{subsec:model},
we present the definition of the generalized susceptibility calculated by RPA in Sec.~\ref{subsec:method-susceptibility}.
We describe how to analyze the eigenmodes of the susceptibility and classify the fluctuations in Sec.~\ref{subsec:method-eigenmode-analysis}.
In Sec.~\ref{sec:result}, we show the results on the electronic structure and fluctuations enhanced by electron correlations for AlV$_2$O$_4$ and LiV$_2$O$_4$.
After presenting the band structures and the tight-binding parameters in Sec.~\ref{subsec:result-band}, we show the results by the eigenmode analyses of the generalized susceptibilities in Sec.~\ref{subsec:result-susceptibility}.
The results are discussed in detail in Sec.~\ref{sec:discussion}.
Section~\ref{sec:summary} is devoted to summary.

\section{Model and Method\label{sec:model-method}}

\subsection{Multiband Hubbard model\label{subsec:model}}
In order to estimate realistic model parameters for each compound, we start with {\it ab initio} calculations based on the density functional theory with the local density approximation (LDA)~\cite{P.Hohenberg-1964-LDA, W.Kohn-1965-LDA}.
In the LDA calculation, we use a fully-relativistic first-principles computational code, QMAS~\cite{QMAS}.
See Ref.~\onlinecite{Shinaoka-2012-Cd2Os2O7} for the details of {\it ab initio} calculations.

We construct the multiband Hubbard models by using the MLWFs, which are obtained from the LDA band structures.
In the present study for AlV$_2$O$_4$ and LiV$_2$O$_4$, $t_{\rm 2g}$ bands are energetically separated from others (see Fig.~\ref{fig:Band}), and hence, we adopt the models for electrons in the $t_{\rm 2g}$ orbitals, similarly to our previous study for Cd$_2$Os$_2$O$_7$~\cite{A.Uehara-Cd2Os2O7-2015}.
The Hamiltonian consists of two parts, the one-body part $H_0$ and two-body interaction part $H_1$ as
\begin{align}
H = H_0 + H_1. \label{eq:H0H1}
\end{align}
The one-body part consists of three terms:
\begin{align}
H_0 = H_{t} + H_{\rm trig} + H_{\rm SOI}, \label{eq:H0}
\end{align}
where $H_t$ denotes the kinetic energy of electrons, $H_{\rm trig}$ the trigonal crystal field splitting, and $H_{\rm SOI}$ the relativistic spin-orbit interaction.
Meanwhile, the two-body interaction part consists of two terms:
the Coulomb interactions acting for electrons at the same atomic site and those between different sites, as
\begin{align}
H_1 = H_{1}^{\rm on-site} + H_{1}^{\rm inter-site}. \label{eq:H1}
\end{align}

We estimate the one-body parameters in $H_0$ by using the MLWFs as follows.
The kinetic term $H_t$ in Eq.~\eqref{eq:H0} is given in the form
\begin{align}
H_t &= \sum_{\bm{R}, \rho, \bm{R}', \rho'} \sum_{\zeta,\zeta'}
t_{\zeta,\zeta'} \left[ ( \bm{R} + \bm{r}_{\rho}) - ( \bm{R}' + \bm{r}_{\rho'})\right] \nonumber\\
&\times
\sum_{\sigma}
c_{\zeta  \sigma \bm{R} +\bm{r}_{\rho} }^{\dagger}
c_{\zeta' \sigma \bm{R}'+\bm{r}_{\rho'}}^{\;},
\label{eq:def-transfer}
\end{align}
where $\bm{R}$, $\zeta$, $\sigma$, and $\rho$ denote the unit cell, orbital, spin, and sublattice indices, respectively;
$\bm{r}_{\rho}$ denotes the position vector of the sublattice $\rho$ in the unit cell.
$c^\dagger$ denotes the creation operator for the MLWF.
The transfer integral $t_{\zeta,\zeta'}$ is given by the overlap integral between the MLWFs.
On the other hand, the trigonal crystal field for the $t_{\rm 2g}$ manifold takes the form of
\begin{align}
H_{\mathrm{trig}}
&= \Delta_{\rm trig} \times
\begin{pmatrix}
0 & 1 & 1 \cr
1 & 0 & 1 \cr
1 & 1 & 0
\end{pmatrix},
\label{eq:HtrigT2gMatrix}
\end{align}
where the basis is taken as $d_{yz}$, $d_{zx}$, and $d_{xy}$.
Besides, the relativistic spin-orbit interaction is given as
\begin{align}
H_{\mathrm{SOI} } &= \lambda_{\rm SOI} \bm{l} \cdot \bm{s}
=
\frac{\lambda_{\rm SOI}}{2} \times
\begin{pmatrix}
 0         & -i\sigma_z &  i\sigma_y \cr
 i\sigma_z &  0         & -i\sigma_x \cr
-i\sigma_y &  i\sigma_x &  0
\end{pmatrix},
\label{eq:HSOIMatrix}
\end{align}
where the basis is taken as $(d_{yz} \uparrow)$, $(d_{yz} \downarrow)$, $(d_{zx} \uparrow)$, $(d_{zx} \downarrow)$, $(d_{xy} \uparrow)$, and $(d_{xy} \downarrow)$;
$\sigma_{x}$, $\sigma_{y}$, and $\sigma_{z}$ denote the Pauli matrices for spin indices.
To estimate the coupling constants, we use the following relations
\begin{align}
\Delta_{\rm trig} &= \int d\bm{r}
\psi^*_{\zeta \sigma}(\bm{r})
H^{\rm LDA}(\bm{r})
\psi_{\zeta' \sigma}(\bm{r})
\quad(\zeta \neq \zeta'),
\\
\frac{\lambda_{\rm SOI}}{2} &= \int d\bm{r}
\psi^*_{xy \uparrow}(\bm{r})
H^{\rm LDA}(\bm{r})
\psi_{yz \downarrow}(\bm{r}),
\end{align}
where $\psi_{\zeta\sigma}(\bm{r})$ is the MLWF and $H^{\rm LDA}(\bm{r})$ is the Hamiltonian in the Kohn-Sham equation.

Next, we introduce the two-body interaction parts.
The on-site part of the electron-electron interaction $H_1^{\rm on-site}$ in Eq.~\eqref{eq:H1} is given by
\begin{align}
H_1^{\rm on-site} &= \frac12 \sum_{\alpha\beta;\alpha'\beta'} U_{\alpha\beta;\alpha'\beta'}\nonumber\\
&\times\sum_{\bm{R}\rho}
c_{\alpha \bm{R}+\bm{r}_{\rho}}^{\dagger}
c_{\beta  \bm{R}+\bm{r}_{\rho}}^{\dagger}
c_{\beta' \bm{R}+\bm{r}_{\rho}}^{\;}
c_{\alpha'\bm{R}+\bm{r}_{\rho}}^{\;},
\label{eq:def-H1-onsite}
\end{align}
where $\alpha$ denotes the orbital and spin indices as $\alpha = (\zeta_{\alpha}, \sigma_{\alpha})$.
Assuming the rotational symmetry of the Coulomb interaction, we take the coupling constant in the form of
\begin{align}
U_{\zeta\mu,\zeta'\mu'} = (U-2J_{\rm H}) \delta_{\zeta \zeta'} \delta_{\mu \mu'}
+ J_{\rm H} (\delta_{\zeta \mu} \delta_{\zeta' \mu'} + \delta_{\zeta \mu'} \delta_{\mu \zeta'} ),
\end{align}
where $U$ and $J_{\rm H}$ denote the Coulomb interaction between the same orbitals and the Hund's-rule coupling, respectively, and $\delta_{\alpha\beta} \equiv \delta_{\zeta_{\alpha}\zeta_{\beta}} \delta_{\sigma_{\alpha}\sigma_{\beta}}$ denotes the Kronecker delta.

In addition to the on-site part, we introduce the inter-site part of the electron interaction, $H_1^{\rm inter-site}$.
Although the Coulomb interaction is long-ranged, for simplicity, we take into account only the dominant component between nearest-neighbor sites:
\begin{align}
H_1^{\rm inter-site}
= V \sum_{\langle \bm{r}, \bm{r}' \rangle } n_{\bm{r}} n_{\bm{r}'}
= \frac{V}{2} \sum_{\bm{r} \bm{\xi}} n_{\bm{r}} n_{\bm{r}+\bm{\xi}},
\label{eq:def-H1-intersite}
\end{align}
where $n_{\bm{r}} = \sum_{\alpha} c_{\alpha \bm{r}}^{\dagger} c_{\alpha \bm{r}}^{\;}$ is the local electron density at site $\bm{r}$, and
$\bm{\xi}$ denotes the vector connecting the nearest-neighbor sites.

The two-body interactions are incorporated at the level of RPA, as described in the next section.
In the RPA calculations, we treat the coupling constants, $U$, $J_{\rm H}$, and $V$ as parameters.

\subsection{Generalized susceptibility\label{subsec:method-susceptibility}}
On the basis of the one-body Hamiltonian, $H_0$, we first calculate the static bare susceptibility $\chi^{(0)}_{\alpha\beta\rho;\alpha'\beta'\rho'}(\bm{q})$.
The definition is given in the form
\begin{align}
\chi^{(0)}_{\alpha\beta\rho;\alpha'\beta'\rho'}(\bm{q}) =
- T
\int d\bm{k} \sum_{\omega_k}
\mathcal{G}^{(0)}_{\alpha'\rho';\alpha \rho }(k)
\mathcal{G}^{(0)}_{\beta  \rho ;\beta' \rho'}(k+q),
\end{align}
where $\mathcal{G}^{(0)}_{\alpha'\rho';\alpha \rho}(k)$ is the non-interacting Green function.
$k$ is the four dimensional wave vector $k=(\bm{k}, \omega_k)$ where $\bm{k}$ and $\omega_k$ denotes the wave number vector and Matsubara frequency, respectively.
Here, the matrix $\chi^{(0)}$ is defined for the orbital and spin indices, $\alpha$ and $\beta$ [$\alpha = (\zeta_{\alpha}, \sigma_{\alpha})$; $\zeta_\alpha$ and $\sigma_\alpha$ denote the orbital and spin, respectively], and the sublattice indices, $\rho$.

Next, we calculate the generalized susceptibility by including the effect of electron correlations in a perturbative way at the level of RPA.
For the on-site interaction $H_1^{\rm on-site}$ in Eq.~(\ref{eq:def-H1-onsite}), the RPA vertex function is defined by
\begin{align}
\mathcal{U}_{ \alpha_1 \beta_1 \rho_1; \alpha_2 \beta_2 \rho_2}
\equiv
\delta_{\rho_1 \rho_2} (U_{\alpha_1 \beta_1; \beta_2 \alpha_2} + U_{\alpha_1 \beta_2; \alpha_2 \beta_1}),
\label{eq:U-interaction-matrix}
\end{align}
where the first and second terms correspond to the so-called bubble and ladder contributions, respectively.
The Dyson equation is written in the form of
\begin{align}
\chi^{\rm RPA}_{l;l'}(\bm{q})
&=
\chi^{(0)}_{l;l'}(\bm{q}) + \sum_{l_1 l_2} \chi^{(0)}_{l;l_1}(\bm{q}) \mathcal{U}_{l_1;l_2} \chi^{\rm RPA}_{l_2;l'}(\bm{q}),
\end{align}
where $l$ denotes the set of ($\alpha, \beta, \rho$).
This gives $\chi^{\rm RPA}$ in the matrix form as
\begin{align}
\chi^{\rm RPA}(\bm{q}) =
\left[I - \chi^{(0)}(\bm{q}) \mathcal{U} \right]^{-1} \chi^{(0)}(\bm{q}),
\label{eq:U-Dyson-matrix}
\end{align}
where $I$ denotes the unit matrix.

On the other hand, for the inter-site interaction in Eq.~(\ref{eq:def-H1-intersite}), the mode coupling appears in the ladder contribution, which is not easy to handle in the perturbation.
Hence, we omit the ladder contribution in $\chi^{\rm RPA}$ when including the inter-site electron interaction.
Consequently, $\chi^{\rm RPA}$ for this case is given in the same form as Eq.~\eqref{eq:U-Dyson-matrix} while replacing Eq.~\eqref{eq:U-interaction-matrix} by
\begin{align}
\mathcal{U}_{ \alpha_1 \beta_1 \rho_1 ; \beta_2 \alpha_2 \rho_2}(\bm{q}) &\equiv
  U_{\alpha_1 \beta_1;\beta_2\alpha_2} \delta_{\rho_1 \rho_2} +
  V_{\rho_1 \rho_2}( \bm{q}) \delta_{\alpha_1 \alpha_2} \delta_{\beta_1 \beta_2},
\end{align}
where $V_{\rho_1 \rho_2}( \bm{q})$ is the coupling constant of the inter-site interaction in the Fourier transformed representation.

We note that the combined method of LDA and RPA potentially has so-called double-counting problem of electron correlations.
In the present study, however, we do not introduce the double-counting correction for the following reason.
In similar theoretical schemes, such as LDA+DMFT~\cite{M.Karolak-LDA-DMFT-double-counting}, the double-counting correction is introduced for adjusting the energy gap between $d$ and $p$ levels when the electron correlations are taken into account only for $d$ electrons in the effective model.
Meanwhile, when the model deduced from LDA includes only the $d$ levels, the double-counting correction is not relevant.
In the present calculations, we construct the multiband Hubbard model only for the $d$ orbitals, thus we do not consider the double-counting correction.

\subsection{Eigenmode analysis\label{subsec:method-eigenmode-analysis}}
From the generalized susceptibility $\chi^{\rm RPA}$, we can extract the information of fluctuations in the multiple degrees of freedom as follows.
According to the fluctuation dissipation theorem, the generalized susceptibility satisfies the following relation
\begin{align}
\delta \langle c_{\alpha\rho\bm{q}}^{\dagger} c_{\beta\rho\bm{q}}^{\;} \rangle
=
\sum_{\alpha'\beta'\rho'} \chi^{\rm RPA}_{\alpha\beta\rho ;\alpha'\beta'\rho'}(\bm{q}) h_{\alpha'\beta'\rho'}(\bm{q}).
\label{eq:fluctuation-dissipation}
\end{align}
Here, the Fourier transform of the creation operator is given by
$c_{\alpha \rho \bm{q}}^{\dagger} \equiv (1/\Omega) \sum_{\bm{R}} e^{i\bm{q}\cdot ( \bm{R}+\bm{r}_{\rho})} c_{\alpha \bm{R} + \bm{r}_{\rho}}^{\dagger}$ where $\Omega$ denotes the system size;
$h_{\alpha\beta\rho}$ denotes a generalized external field conjugate to $c_{\alpha\rho\bm{q}}^{\dagger} c_{\beta\rho\bm{q}}^{\;}$, and
$\delta \langle A\rangle$ represents the difference of the thermal average of $A$ from that in the absence of the external field.
If the external field is parallel to the $\kappa$th eigenvector $e_{\alpha\beta\rho}^{\kappa}$ of $\chi^{\rm RPA}$, namely,
$h_{\alpha\beta\rho}^{\kappa}(\bm{q}) \propto e^{\kappa}_{\alpha\beta\rho}(\bm{q})$,
Eq.~\eqref{eq:fluctuation-dissipation} is rewritten into the form of
\begin{align}
\delta \langle c_{\alpha\rho\bm{q}}^{\dagger} c_{\beta\rho\bm{q}}^{\;} \rangle = x^{\kappa}(\bm{q}) h_{\alpha\beta\rho}^{\kappa}(\bm{q}),
\label{eq:fluctuation-dissipation-parallel}
\end{align}
where $x^{\kappa}(\bm{q})$ is the corresponding eigenvalue.
This indicates that $\delta \langle c_{\alpha\rho\bm{q}}^{\dagger} c_{\beta\rho\bm{q}}^{\;} \rangle$ also becomes parallel to $e_{\alpha\beta\rho}^{\kappa}$.
Thus, the eigenvector is regarded as the direction of the fluctuation $\delta \langle c_{\alpha\rho\bm{q}}^{\dagger} c_{\beta\rho\bm{q}}^{\;} \rangle$, and the eigenvalue $x^{\kappa}(\bm{q})$ corresponds to the amplitude of the fluctuation.
Thus, the analysis of the eigenvalues and eigenvectors provides the information of fluctuations.

Specifically, the charge fluctuation $\delta n_{\rho}(\bm{q})$ at the sublattice $\rho$ is defined by
\begin{align}
\delta n_{\rho}(\bm{q})
= \delta \left\langle \sum_{\alpha} c_{\alpha\rho\bm{q}}^{\dagger} c_{\alpha\rho\bm{q}}^{\;} \right\rangle
\propto \sum_{\alpha} e_{\alpha \alpha \rho}^{\kappa}(\bm{q}).
\end{align}
$\delta n_{\rho}(\bm{q})$ can be decomposed into the orbital components as
\begin{align}
\delta n_{\rho}(\bm{q}) = \sum_{\zeta} \delta n_{\zeta \rho}(\bm{q}),
\label{eq:fluctuation-delta-n}
\end{align}
where $\zeta$ denotes the orbital index.
Here,
\begin{align}
\delta n_{\zeta \rho}(\bm{q})
= \delta \left\langle \sum_{\sigma} c_{\zeta \sigma \rho\bm{q}}^{\dagger} c_{\zeta \sigma \rho\bm{q}}^{\;} \right\rangle
\propto \sum_{\sigma} e_{(\zeta, \sigma) (\zeta, \sigma) \rho}(\bm{q}),
\label{eq:fluctuation-delta-n-orbital}
\end{align}
where $\sigma$ denotes the spin index.
Similarly, the spin fluctuation $\delta \bm{s}_{\rho}(\bm{q})$ is defined by
\begin{align}
\delta \bm{s}_{\rho}(\bm{q})
&=\delta \left\langle \sum_{\zeta \sigma\sigma'} c_{\zeta \sigma \rho\bm{q}}^{\dagger} \bm{\sigma}_{\sigma\sigma'}c_{\zeta \sigma' \rho\bm{q}}^{\;} \right\rangle
\nonumber\\
&\propto\sum_{\zeta \sigma\sigma'} \bm{\sigma}_{\sigma\sigma'} e_{(\zeta, \sigma) (\zeta \sigma') \rho}(\bm{q}),
\label{eq:fluctuation-delta-spin}
\end{align}
where $\bm{\sigma}$ denotes the Pauli matrix.

Finally, let us comment on the way to classify the fluctuation modes into charge and spin components.
We classify the eigenmodes according to the spin dependence as follows.
The charge fluctuation $\delta n$ satisfies the relations
\begin{align}
\delta \langle c_{  \uparrow}^{\dagger}c_{  \uparrow}^{\;} \rangle = \delta \langle c_{\downarrow}^{\dagger}c_{\downarrow}^{\;} \rangle \neq 0, \quad
\delta \langle c_{  \uparrow}^{\dagger}c_{\downarrow}^{\;} \rangle = \delta \langle c_{\downarrow}^{\dagger}c_{  \uparrow}^{\;} \rangle =    0.
\end{align}
On the other hand, the $z$ component of the spin fluctuations satisfies
\begin{align}
\delta \langle c_{  \uparrow}^{\dagger}c_{  \uparrow}^{\;} \rangle = -\delta \langle c_{\downarrow}^{\dagger}c_{\downarrow}^{\;} \rangle \neq 0, \quad
\delta \langle c_{  \uparrow}^{\dagger}c_{\downarrow}^{\;} \rangle =  \delta \langle c_{\downarrow}^{\dagger}c_{  \uparrow}^{\;} \rangle =    0,
\end{align}
and the $x$ and $y$ components satisfy
\begin{align}
\delta \langle c_{  \uparrow}^{\dagger}c_{  \uparrow}^{\;} \rangle =    \delta \langle c_{\downarrow}^{\dagger}c_{\downarrow}^{\;} \rangle   = 0, \quad
\delta \langle c_{  \uparrow}^{\dagger}c_{\downarrow}^{\;} \rangle = \pm\delta \langle c_{\downarrow}^{\dagger}c_{  \uparrow}^{\;} \rangle^* \neq 0.
\end{align}
In addition, we categorize the eigenmodes into acoustic and optical ones according to the sublattice dependence.
The acoustic mode satisfies the relation
\begin{align}
\delta \langle c_{\bm{r}_1}^{\dagger}c_{\bm{r}_1}^{\;} \rangle =
\delta \langle c_{\bm{r}_2}^{\dagger}c_{\bm{r}_2}^{\;} \rangle =
\delta \langle c_{\bm{r}_3}^{\dagger}c_{\bm{r}_3}^{\;} \rangle =
\delta \langle c_{\bm{r}_4}^{\dagger}c_{\bm{r}_4}^{\;} \rangle,
\end{align}
where $\bm{r}_\rho$ denotes the position of the sublattice $\rho$.
On the other hand, the optical mode satisfies the relation
\begin{align}
\sum_{\rho=1}^4 \delta \langle c_{\bm{r}_\rho}^{\dagger}c_{\bm{r}_\rho}^{\;} \rangle = 0.
\end{align}

\section{Result\label{sec:result}}
In this section, we present the results for AlV$_2$O$_4$ and LiV$_2$O$_4$ obtained by the method in Sec.~\ref{sec:model-method}.
In Sec.~\ref{subsec:result-band}, we show the LDA band structures and the tight-binding parameters.
In Sec.~\ref{subsec:result-susceptibility}, we show the generalized susceptibilities with the identification of the dominant fluctuations enhanced by electron correlations.

\subsection{LDA Band structure and MLWFs\label{subsec:result-band}}

\subsubsection{Band structure}
We begin with the LDA calculations of the electronic band structures.
We adopt a face-centered cubic primitive unit cell containing four V ions forming a pyrochlore lattice included in AlV$_2$O$_4$ and LiV$_2$O$_4$.
The number of $k$ points sampled is $4\times 4\times 4$.
MLWFs are also composed by using the QMAS code.

\begin{figure}[thbp]
  \centering
  \includegraphics[width=1.\columnwidth]{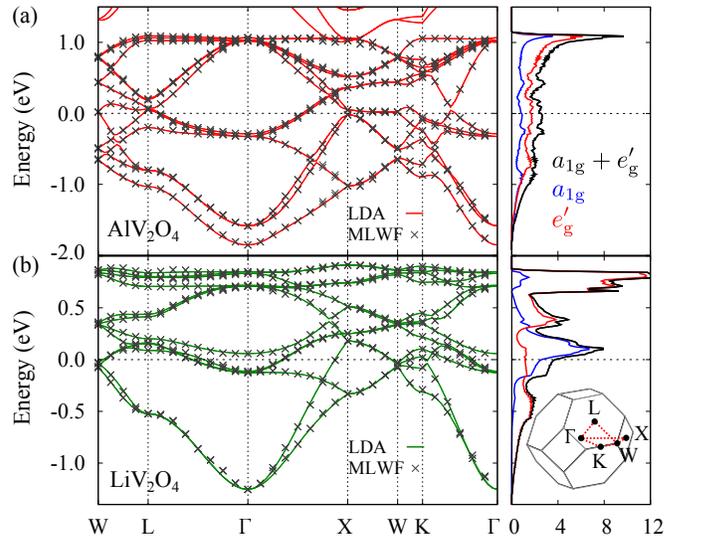}
  \caption{\label{fig:Band}
    (color online).
    Electronic states for (a) AlV$_2$O$_4$ and (b) LiV$_2$O$_4$.
    The Fermi level is set to zero.
    In the left panels, the curves denote the band structures obtained by the relativistic LDA calculations, while the results of the MLWF fitting are shown by the crosses.
    The right panels represent the DOS calculated by the MLWFs.
    The results decomposed into the $a_{\rm 1g}$ and $e'_{\rm g}$ components are also shown.
    The inset in the right panel of (b) represents the first Brillouin zone for the pyrochlore lattice structure composed of V cations.
  }
\end{figure}

The left panels of Fig.~\ref{fig:Band} show the LDA band structures for AlV$_2$O$_4$ and LiV$_2$O$_4$.
The results are obtained for the experimental lattice structures at $T=800$~K for AlV$_2$O$_4$~\cite{T.Katsufuji-private-communication} and 4~K for LiV$_2$O$_4$~\cite{O.Chmaissem-LiV2O4-1997}, respectively.
The result for LiV$_2$O$_4$ well agrees with that in the previous study~\cite{J.Matsuno-LiV2O4-2000}.
In both cases, the relevant bands near the Fermi level are composed of the $3d$ $t_{\rm 2g}$ orbitals of V cations, energetically separated from other bands.
We then perform the MLWF fitting for the $t_{\rm 2g}$ bands~\cite{Marzari-1997-MLWF, Souza-2001-MLWF}.
As shown in Fig.~\ref{fig:Band}, the obtained MLWFs well reproduce the LDA band structure.
From the MLWFs, we construct the non-interacting part of the multiband Hubbard Hamiltonian for each compound, which includes the electron transfer, trigonal crystal field splitting, and spin-orbit coupling.

The density of states (DOS) is shown in the right panels of Fig.~\ref{fig:Band}.
The results for AlV$_2$O$_4$ and LiV$_2$O$_4$ show contrasting behavior.
In AlV$_2$O$_4$, the DOS is featureless near the Fermi level, and the $a_{\rm 1g}$ and two $e'_{\rm g}$ orbitals are almost equally occupied.
On the other hand, in LiV$_2$O$_4$, the DOS near the Fermi level strongly depends on the energy, which is dominated by the $a_{\rm 1g}$ component.
The difference of the DOS mainly comes from the difference in the transfer integrals.

\subsubsection{Tight-binding parameter}
\begin{figure}[thbp]
  \centering
  \includegraphics[width=.5\columnwidth]{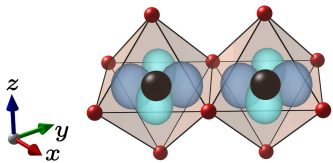}
  \caption{\label{fig:dd-sigma}
    (color online).
    Schematic picture of edge-sharing VO$_6$ octahedra.
    The black and red spheres denote V and O, respectively.
    A pair of $d_{xy}$ orbitals is drawn to show the $dd\sigma$ overlap.
  }
\end{figure}

Tables~\ref{tab:AlV2O4-hopping-matrix} and \ref{tab:LiV2O4-hopping-matrix} show the transfer integrals $t_{\zeta,\zeta'}$ calculated by the MLWFs.
For both compounds, the nearest-neighbor transfer is larger than that of the second and third neighbors.
It is also common that the most dominant transfer is the $dd\sigma$ component between nearest-neighbor sites, which is schematically shown in Fig.~\ref{fig:dd-sigma}.
However, the ratio of the diagonal $dd\sigma$ component to the off-diagonal ones is largely different between the two compounds:
the $dd\sigma$ transfer is more than 30 times as large as the off-diagonal $d_{xy}$-$d_{yz}$ one in AlV$_2$O$_4$, whereas it is about twice of the $d_{zx}$-$d_{yz}$ one in LiV$_2$O$_4$.

\begin{table}[thbp]
  \caption{\label{tab:AlV2O4-hopping-matrix}
    Transfer integrals $t_{\zeta,\zeta'}[\bm{\delta}]$ obtained from the MLWF analysis for AlV$_2$O$_4$ between
    (a) nearest-neighbor,
    (b) second-neighbor, and
    (c) third-neighbor sites in the basis of ($d_{xy}, d_{yz}, d_{zx}$).
    (d) represents the nearest-neighbor transfer integrals in the basis of ($a_{\rm 1g}, e'_{\rm g,1}, e'_{\rm g,2}$).
    The displacement vector $\bm{\delta}$ is ($a/4, a/4, 0$), ($a/2, a/4, -a/4$), ($a/2, a/2, 0$), ($a/4, a/4, 0$) for (a), (b), (c), and (d), respectively.
    $a$ is the lattice constant for the cubic unit cell of the pyrochlore lattice.
    The unit is meV.
    The values less than 1~meV are omitted.
  }
  \begin{minipage}[]{.50\columnwidth}
    \begin{flushleft}
      \begin{tabular}{r|rrr}
        (a)
                 & $d_{xy}$ & $d_{yz}$ & $d_{zx}$ \\ \hline
        $d_{xy}$ &-467      &  15      &  15      \\
        $d_{yz}$ & -15      & 132      &   2      \\
        $d_{zx}$ & -15      &   2      & 132
      \end{tabular}
    \end{flushleft}
  \end{minipage}
  \begin{minipage}[]{.40\columnwidth}
    \begin{flushleft}
      \begin{tabular}{r|rrr}
        (b)
                 & $d_{xy}$ & $d_{yz}$ & $d_{zx}$ \\ \hline
        $d_{xy}$ & 14       & 2        & -9       \\
        $d_{yz}$ & -2       & 4        & -2       \\
        $d_{zx}$ & 11       & 2        & 14
      \end{tabular}
    \end{flushleft}
  \end{minipage}
  \begin{minipage}[]{\columnwidth}
    \hspace{0pt}\\
  \end{minipage}
  \begin{minipage}[]{.50\columnwidth}
    \begin{flushleft}
      \begin{tabular}{r|rrr}
        (c)
                 & $d_{xy}$ & $d_{yz}$ & $d_{zx}$ \\ \hline
        $d_{xy}$ &  -        &  2      &  2       \\
        $d_{yz}$ &  2        &  -      &  8       \\
        $d_{zx}$ &  2        &  8      &  -
      \end{tabular}
    \end{flushleft}
  \end{minipage}
  \begin{minipage}[]{.40\columnwidth}
    \begin{flushleft}
      \begin{tabular}{r|rrr}
        (d)
                       & $a_{\rm 1g}$ & $e'_{\rm g,1}$ & $e'_{\rm g,2}$ \\ \hline
        $a_{\rm 1g}$   &  -66      &  215    &  215     \\
        $e'_{\rm g,1}$ & -185      &  266    &    0     \\
        $e'_{\rm g,2}$ & -185      &    0    & -130
      \end{tabular}
    \end{flushleft}
  \end{minipage}

\end{table}
\begin{table}[thbp]
  \caption{\label{tab:LiV2O4-hopping-matrix}
    Transfer integrals for LiV$_2$O$_4$.
    The notations are the same as in Table~\ref{tab:AlV2O4-hopping-matrix}.
  }
  \begin{minipage}[]{.50\columnwidth}
    \begin{flushleft}
      \begin{tabular}{r|rrr}
        (a)
                 & $d_{xy}$ & $d_{yz}$ & $d_{zx}$ \\ \hline
        $d_{xy}$ &-223      & -45      & -45      \\
        $d_{yz}$ &  45      &  91      &-112      \\
        $d_{zx}$ &  45      &-112      &  91
      \end{tabular}
    \end{flushleft}
  \end{minipage}
  \begin{minipage}[]{.40\columnwidth}
    \begin{flushleft}
      \begin{tabular}{r|rrr}
        (b)
                 & $d_{xy}$ & $d_{yz}$ & $d_{zx}$ \\ \hline
        $d_{xy}$ & 10       & 7        & 4        \\
        $d_{yz}$ & -3       & 1        & -7       \\
        $d_{zx}$ & -2       & 3        & 10
      \end{tabular}
    \end{flushleft}
  \end{minipage}
    \begin{minipage}[]{\columnwidth}
      \hspace{0pt}\\
    \end{minipage}
  \begin{minipage}[]{.50\columnwidth}
    \begin{flushleft}
      \begin{tabular}{r|rrr}
        (c)
                 & $d_{xy}$ & $d_{yz}$ & $d_{zx}$ \\ \hline
        $d_{xy}$ & -56       &  -      &  -       \\
        $d_{yz}$ &  -        &  -      &  5       \\
        $d_{zx}$ &  -        &  5      &  -
      \end{tabular}
    \end{flushleft}
  \end{minipage}
  \begin{minipage}[]{.40\columnwidth}
    \begin{flushleft}
      \begin{tabular}{r|rrr}
        (d)
                       & $a_{\rm 1g}$ & $e'_{\rm g,1}$ & $e'_{\rm g,2}$ \\ \hline
        $a_{\rm 1g}$   &  -88      &   22    &   22     \\
        $e'_{\rm g,1}$ & -112      &  156    &    0     \\
        $e'_{\rm g,2}$ & -112      &    0    & -202
      \end{tabular}
    \end{flushleft}
  \end{minipage}
\end{table}

The difference of the transfer integrals between the two compounds is also seen in the ($a_{\rm 1g}$, $e'_{\rm g}$) representation.
For AlV$_2$O$_4$, as shown in Table~\ref{tab:AlV2O4-hopping-matrix}(d), both of the off-diagonal components are nearly three times as large as the diagonal $a_{\rm 1g}$-$a_{\rm 1g}$ one in magnitude, and comparable to the diagonal $e'_{\rm g}$-$e'_{\rm g}$ ones.
In contrast, for LiV$_2$O$_4$, the diagonal $e'_{\rm g}$-$e'_{\rm g}$ components are dominant, as shown in Table~\ref{tab:LiV2O4-hopping-matrix}(d).
The quantitative difference plays a decisive role in the contrasting behavior between the two compounds, as detailed below.

Unlike the transfer integrals, the coupling constants of the one-body on-site terms are almost the same for the two compounds.
For AlV$_2$O$_4$ and LiV$_2$O$_4$, we obtain the estimates the trigonal crystal field splitting $\Delta_{\rm trig}=53$~meV and 61~meV, respectively, while spin-orbit coupling $\lambda_{\rm SOI}=26$~meV and 27~meV, respectively.

\subsection{Generalized Susceptibility and Eigenmode Analysis\label{subsec:result-susceptibility}}
To study the fluctuations of AlV$_2$O$_4$ and LiV$_2$O$_4$, generalized susceptibility is calculated.
After the computation of the static bare susceptibility, electron correlation is included at the level of RPA.
By the eigenmode analysis of the generalized susceptibility, charge, spin, and orbital fluctuations are obtained.

\subsubsection{Bare susceptibility}
Based on the non-interacting Hamiltonian constructed from the MLWFs, we first calculate the static bare susceptibility $\chi^{(0)}_{\alpha\beta\rho;\alpha'\beta'\rho'}(\bm{q})$.
The matrix size is $144\times 144$: four sublattices (only diagonal) with three orbital and two spin components [$4\times (3\times 2)^2 = 144$].
In the calculations, we replace the integral of $\bm{k}$ over the first Brillouin zone by the summation for $32^3$ $\bm{k}$ grid points, while we take the summation of $\omega_k$ analytically.

\begin{figure}[thbp]
  \centering
  \includegraphics[width=1.\columnwidth]{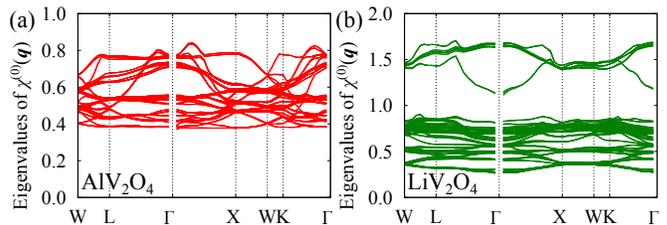}
  \caption{\label{fig:chi0}
    (color online).
    Eigenvalues of $\chi^{(0)}(\bm{q})$ calculated by using the tight-binding parameters estimated by the MLWFs for (a) AlV$_2$O$_4$ and (b) LiV$_2$O$_4$.
    The calculations are done at $T=10$~K.
  }
\end{figure}

Figure~\ref{fig:chi0} shows all the eigenvalues of $\chi^{(0)}(\bm{q})$ for AlV$_2$O$_4$ and LiV$_2$O$_4$.
The results look very differently: the eigenmodes are rather entangled in the entire spectrum for AlV$_2$O$_4$, whereas the result for LiV$_2$O$_4$ shows sixteen eigenmodes well separated from the other modes.
Analyzing the eigenvectors of the sixteen eigenmodes, we find that they are dominantly in the $a_{\rm 1g}$ orbital sector, which is anticipated from the DOS in Fig.~\ref{fig:Band}(b).

\subsubsection{RPA calculation and eigenmode analysis of AlV$_2$O$_4$}
Next, we calculate the generalized susceptibility by including the effect of electron correlations in a perturbative way at the level of RPA.
First, let us discuss the results for AlV$_2$O$_4$.
We investigate the generalized susceptibility obtained by RPA, $\chi^{\rm RPA}$, while changing $U$, $J_{\rm H}$, and $V$.
Among the parameters, we find that $V$ induces particularly interesting behavior, which might be related with the heptamer formation in AlV$_2$O$_4$, as discussed below.

\begin{figure}[thbp]
  \centering
  \includegraphics[width=1.\columnwidth]{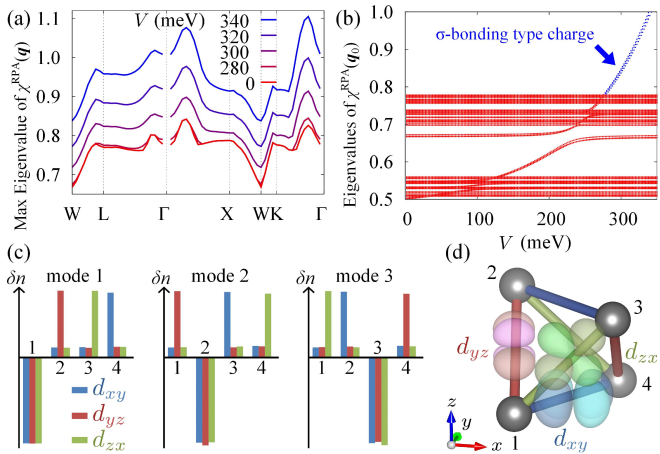}
  \caption{\label{fig:chiRPA-AlV2O4}
    (color online).
    (a) $V$ dependence of the maximum eigenvalues of $\chi^{\rm RPA}(\bm{q})$ for AlV$_2$O$_4$.
    (b) $V$ dependence of the eigenvalues at $\bm{q}_0 = (\pi/16, \pi/16, \pi/16)$, which is the smallest wave number along the $\Gamma$-L line in the present calculations.
    In (a) and (b), we take $U=300$~meV and $J_{\rm H}=30$~meV, while the other parameters are the same as in Fig.~\ref{fig:chi0}(a).
    (c) $\sigma$-bonding type charge fluctuations obtained from the eigenmode analysis.
    The vertical axis represents the fluctuation of the electron density in arbitrary units.
    The histogram represents the density fluctuations decomposed into the $d_{xy}$, $d_{yz}$, and $d_{zx}$ orbitals.
    The numbers 1-4 in the horizontal axis denote the sublattices.
    (d) Schematic visualization of the fluctuations in the mode 1 in (a).
  }
\end{figure}

Figure~\ref{fig:chiRPA-AlV2O4}(a) shows the largest eigenvalues of $\chi^{\rm RPA}$, which are considerably enhanced by increasing $V$.
As shown in Fig.~\ref{fig:chiRPA-AlV2O4}(b), the enhancement occurs in particular eigenmodes, whereas all the other modes are almost insensitive to $V$.
To clarify the nature of these enhanced fluctuations by $V$, we analyze the eigenvectors of the three quasi-degenerate modes~\cite{A.Uehara-Cd2Os2O7-2015}.
We find that the dominant fluctuations are in the charge sector.
Figure \ref{fig:chiRPA-AlV2O4}(c) shows the fluctuations of local electron densities.
In all the three modes, the density fluctuation at one sublattice has the opposite sign to the other three, and the net density fluctuation vanishes in the four-site tetrahedron.
Note that, while the densities at the sublattices 1, 2, and 3 are suppressed in the modes 1, 2, and 3, respectively,
one can make the mode 4 in which the sublattice 4 is suppressed by a linear combination of the modes 1-3.

Interestingly, we find that the density fluctuations in these modes are strongly orbital dependent, as shown in Fig.~\ref{fig:chiRPA-AlV2O4}(c).
The orbital dependence indicates that the dominant fluctuations occur through the $dd\sigma$ orbital on each bond.
For instance, in the mode 1, the charge density in the $d_{yz}$ orbital is dominantly transferred between the sites 1 and 2 on the $yz$ plane, which is regarded as the charge fluctuation through the $dd\sigma$ orbital.
The bond- and orbital-dependent fluctuations are schematically shown in Fig.~\ref{fig:chiRPA-AlV2O4}(d).
Thus, the fluctuations of the three modes sensitive to $V$ are of $\sigma$-bonding type.
The importance of such $\sigma$-bonding states in the heptamer was suggested in the previous experimental and theoretical studies~\cite{Y.Horibe-AlV2O4-2006, K.Matsuda-AlV2O4-2006}.
Hence, our results for the dominant charge fluctuations in $\sigma$-bonding orbitals suggest that the inter-site Coulomb repulsion plays a role in the self-organization of heptamers in AlV$_2$O$_4$.

We note that the values of $V$ in the present RPA calculations are rather large: in reality, the bare value of $V$ will be considerably smaller than $U$.
Nevertheless, our finding might be relevant to the heptamer formation in AlV$_2$O$_4$ due to the following reason.
The structural change associated with the heptamer formation clearly indicates the importance of the Peierls-type electron-phonon coupling.
Such inter-site phonons are known to give rise to an effective repulsive interaction for electrons: indeed, the integration of phonon degrees of freedom leaves the effective $V$ term, together with other inter-site interactions~\cite{A.Heeger-Polymer-1988}.
We regard that such effects are included in the value of $V$ in the RPA analysis, although the realistic estimate is left for future study.

\subsubsection{RPA calculation and eigenmode analysis of LiV$_2$O$_4$}
\begin{figure}[thbp]
  \centering
  \includegraphics[width=1.\columnwidth]{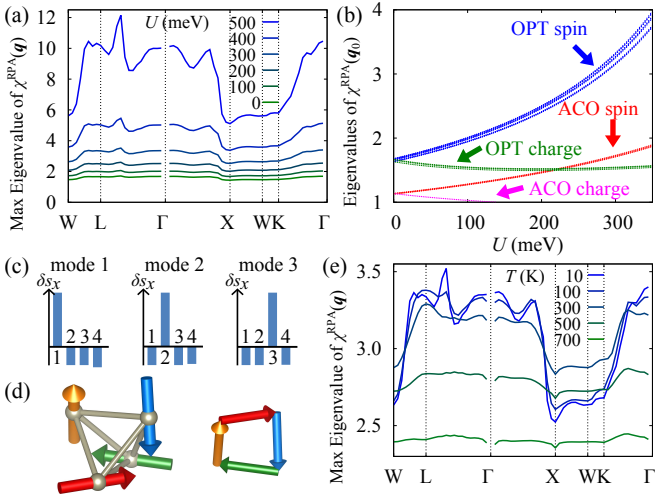}
  \caption{\label{fig:chiRPA-LiV2O4}
    (color online).
    (a) $U$ dependence of the maximum eigenvalues of $\chi^{\rm RPA}(\bm{q})$ for LiV$_2$O$_4$.
    (b) $U$ dependence of the largest sixteen eigenvalues at $\bm{q}_0$.
    OPT (ACO) represents optical (acoustic: uniform in the unit cell) type fluctuation.
    In (a) and (b), the calculations are done at $J_{\rm H}/U=0.1$ and $V=0$, while the other parameters are taken to be the same as in Fig.~\ref{fig:chi0}(b).
    (c) OPT spin fluctuations obtained from the eigenmode analysis.
    The vertical axis represents the spin fluctuations of the $x$ component in arbitrary units.
    The other components are qualitatively similar.
    (d) Schematic visualization of the OPT spin fluctuations, in which the net spin fluctuation vanishes in the four-site tetrahedron.
    (e) $T$ dependence of the maximum eigenvalues of $\chi^{\rm RPA}(\bm{q})$ at $U=300$~meV, $J_{\rm H}=30$~meV, and $V=0$.
  }
\end{figure}

Next, we discuss the results of $\chi^{\rm RPA}$ for LiV$_2$O$_4$.
We here focus on the effect of $U$, while $V$ leads to a different charge fluctuation from AlV$_2$O$_4$ as mentioned below.
Figure \ref{fig:chiRPA-LiV2O4}(a) shows $U$ dependence of the maximum eigenvalues of $\chi^{\rm RPA}$.
The maximum eigenvalues are enhanced by $U$ and become more dispersive.

We plot $U$ dependence of the largest sixteen eigenvalues of $\chi^{\rm RPA}$ in Fig.~\ref{fig:chiRPA-LiV2O4}(b).
Note that all of them are of $a_{\rm 1g}$ character as discussed above.
We find that nine eigenmodes are largely enhanced by $U$.
We also elucidate that all the nine eigenmodes consist of spin fluctuations.
Figure~\ref{fig:chiRPA-LiV2O4}(c) shows the three of them, which are the spin fluctuations in the $x$ component.
In all the three modes, the spin fluctuation $\delta s^x$ at one sublattice has an opposite sign to the other three;
the net $\delta s^x$ vanishes in the four-site tetrahedron.
Similar situations are found also in the $y$ and $z$ components.
Namely, the spin fluctuations satisfy the relation $\sum_{\rho=1}^{4} \delta \bm{s}_{\rho} = 0$, where $\delta\bm{s}_\rho = (\delta s_\rho^x, \delta s_\rho^y, \delta s_\rho^z)$.
Hence, we call them the optical-type spin fluctuations, whose schematic visualization is shown in Fig.~\ref{fig:chiRPA-LiV2O4}(d).
As the zero-sum condition is often met in frustrated spin systems, our results suggest that the dominant fluctuations in LiV$_2$O$_4$ appear in the $a_{\rm 1g}$ spins under strong geometrical frustration.

On the other hand, we find that $V$ enhances a charge fluctuation in the $a_{\rm 1g}$ orbital, where the net $a_{\rm 1g}$ density fluctuation vanishes in the four-site tetrahedron (not shown here).
The result is in sharp contrast to the case of AlV$_2$O$_4$, in which $\sigma$-type fluctuation is enhanced by $V$.

Figure \ref{fig:chiRPA-LiV2O4}(e) shows $T$ dependence of the maximum eigenvalues of $\chi^{\rm RPA}$ for LiV$_2$O$_4$.
The dominant instability is always in the optical-type spin fluctuations in the $a_{\rm 1g}$ orbital.
Although the mode is almost $\bm{q}$ independent at high $T\gtrsim 500$~K, it develops the $\bm{q}$ dependence below $\sim 300$~K.
At lower $T$ below 100~K, the peak between the $\Gamma$-L line develops, being the dominant instability.
We note that a similar peak appeared in the previous theoretical study~\cite{V.Yushankhai-LiV2O4-2007}, while the model and the RPA treatment were different from ours.
The growth of the incommensurate peak is potentially related with that observed in the inelastic neutron scattering for LiV$_2$O$_4$ below 80~K~\cite{S.Lee-LiV2O4-2001, K.Tomiyasu-LiV2O4-2014}.
Nevertheless, further studies beyond the current weak-coupling approach are necessary to address the heavy fermion behavior, as the recent NMR experiment suggests local moments in the $a_{\rm 1g}$ orbital even at high $T$~\cite{Y.Shimizu-LiV2O4-2012}.

\section{Discussion\label{sec:discussion}}
It is worth noting that the electronic structure obtained by LDA and MLWF analyses plays an important role in the dominant fluctuations enhanced by electron correlations.
In AlV$_2$O$_4$, three $t_{\rm 2g}$ orbitals almost equally contribute to the DOS near the Fermi level as shown in Fig.~\ref{fig:Band}(a), and furthermore, the $dd\sigma$ components in the transfer integrals are dominant as shown in Table~\ref{tab:AlV2O4-hopping-matrix}.
These are directly reflected in the dominant $\sigma$-type charge fluctuation in Fig.~\ref{fig:chiRPA-AlV2O4}.
On the other hand, in LiV$_2$O$_4$, the diagonal components represented in the ($a_{\rm 1g}$, $e'_{\rm g}$) basis dominate the transfer integrals, as shown in Table~\ref{tab:LiV2O4-hopping-matrix}(d).
This is closely related with the dominant $a_{\rm 1g}$ spin fluctuation in Fig.~\ref{fig:chiRPA-LiV2O4}.
Thus, for both compounds, the orbital channel for the dominant fluctuations are deduced from the transfer integrals composed from the LDA band structures, although
the actual types of fluctuations are clarified only after performing RPA.

Finally, we discuss the pressure effect on LiV$_2$O$_4$ in the present scheme.
We performed the LDA calculation by reducing the lattice constant by $2$\% from the original value, which approximately corresponds to the situation at the pressure 6.3~GPa~\cite{K.Takeda-LiV2O4-2005}.
By the energy optimization with respect to the $u$ parameter, which controls the trigonal distortion, we find that the value of $u$ decreases from 0.262 to 0.259.
Although it approaches that for AlV$_2$O$_4$ ($u=0.251$)~\cite{T.Katsufuji-private-communication}, the $a_{\rm 1g}$ component remains dominant in the DOS.
Accordingly, our calculations for the generalized susceptibility indicate that the dominant instability is still in the $a_{\rm 1g}$ optical-type spin fluctuations and does not qualitatively change from ambient pressure.
The results imply that the pressure-induced metal-insulator transition in LiV$_2$O$_4$ is caused by a different mechanism from the cluster formation in AlV$_2$O$_4$, and hence, the associated structural change is also likely to be different from AlV$_2$O$_4$.

\section{Summary\label{sec:summary}}
In summary, for a systematic understanding of the contrasting behavior in two mixed valence spinels, AlV$_2$O$_4$ and LiV$_2$O$_4$, we have analyzed charge-spin-orbital fluctuations in the multiband Hubbard models, whose one-body part is constructed by LDA calculations and MLWF analysis.
From the eigenmode analysis of the generalized susceptibility obtained by RPA, we found that, in AlV$_2$O$_4$, the $\sigma$-bonding type charge-orbital fluctuations are enhanced by the intersite repulsion, which provides a clue for understanding of the seven-site cluster formation.
In contrast, for LiV$_2$O$_4$, the optical-type spin fluctuation in the $a_{\rm 1g}$ orbital is strongly enhanced by $U$ at an incommensurate wave number at low temperatures.
We also discussed the pressure effect on LiV$_2$O$_4$ and deduced that the mechanism of pressure-induced metal-insulator transition might be different from the cluster formation in AlV$_2$O$_4$.

As the present analysis is based on the LDA band structures and the electron correlations are included at the level of RPA, it is difficult to discuss the strong correlation effects, for instance, the large mass enhancement at low temperature in LiV$_2$O$_4$.
It is necessary to adopt more sophisticated methods that can treat the renormalization of the band structures and occupations of each orbitals.
This is left for future study.

\section*{Acknowledgement}
We acknowledge fruitful discussions with M.~Itoh, T.~Misawa, Y.~Shimizu, H.~Takeda, K.~Tomiyasu, M.~Udagawa, and Y.~Yamaji.
We thank T.~Katsufuji for providing us the lattice structure data of AlV$_2$O$_4$.
A.U. is supported by Japan Society for the Promotion of Science through Program for Leading Graduate Schools (MERIT).
H.S. acknowledges support from the DFG via FOR 1346, the SNF Grant 200021E-149122, ERC Advanced Grant SIMCOFE and NCCR MARVEL.
Part of the calculations were performed on the Brutus clusters of ETH Z\"{u}rich.
The crystal structures are visualized by using VESTA 3~\cite{VESTA}.


\end{document}